\documentclass[11pt]{article}

\usepackage[english]{babel}
\usepackage[latin1]{inputenc}
\usepackage{graphicx}
\usepackage{color}
\usepackage[hidelinks]{hyperref}
\usepackage{amsmath,amsfonts}
\usepackage{natbib}
\usepackage{bm}
\usepackage{enumerate}
\usepackage{multirow}
\usepackage{booktabs}
\usepackage{subcaption}
\usepackage[flushleft]{threeparttable}

\bibpunct{(}{)}{;}{a}{,}{,}

\title{\Large\textbf{SemiCompRisks: An R Package for Independent and Cluster-Correlated Analyses of Semi-Competing Risks Data}} 
\author{\, \vspace{0.3cm} \\ \textbf{Danilo Alvares}\\Harvard T.H. Chan School of Public Health, Boston, MA, USA \vspace{0.5cm} \\ \textbf{Sebastien Haneuse}\\Harvard T.H. Chan School of Public Health, Boston, MA, USA \vspace{0.5cm} \\ \textbf{Catherine Lee}\\Kaiser Permanente Northern California, Oakland, CA, USA \vspace{0.5cm} \\ \textbf{Kyu Ha Lee}\\The Forsyth Institute, Cambridge, MA, USA}

\date{}

\begin{document} 

\baselineskip18pt

\maketitle 

\begin{abstract}
\noindent Semi-competing risks refer to the setting where primary scientific interest lies in estimation and inference with respect to a non-terminal event, the occurrence of which is subject to a terminal event. In this paper, we present the R package \textbf{SemiCompRisks} that provides functions to perform the analysis of independent/clustered semi-competing risks data under the illness-death multi-state model. The package allows the user to choose the specification for model components from a range of options giving users substantial flexibility, including: accelerated failure time or proportional hazards regression models; parametric or non-parametric specifications for baseline survival functions; parametric or non-parametric specifications for random effects distributions when the data are cluster-correlated; and, a Markov or semi-Markov specification for terminal event following non-terminal event. While estimation is mainly performed within the Bayesian paradigm, the package also provides the maximum likelihood estimation for select parametric models. The package also includes functions for univariate survival analysis as complementary analysis tools. \vspace{0.3cm} 
\end{abstract}

\newpage

\section{Introduction} \label{sec:intro}

Semi-competing risks refer to the general setting where primary scientific interest lies in estimation and inference with respect to a non-terminal event (e.g., disease diagnosis), the occurrence of which is subject to a terminal event (e.g., death) \citep{fine2001, jazic2016}. When there is a strong association between two event times, na\"ive application of a univariate survival model for non-terminal event time will result in overestimation of outcome rates as the analysis treats the terminal event as an independent censoring mechanism \citep{haneuse2016}. The semi-competing risks analysis framework appropriately treats the terminal event as a competing event and considers the dependence between non-terminal and terminal events as part of the model specification.

\begin{figure}[htb]
	\centering
		\includegraphics[width=16cm]{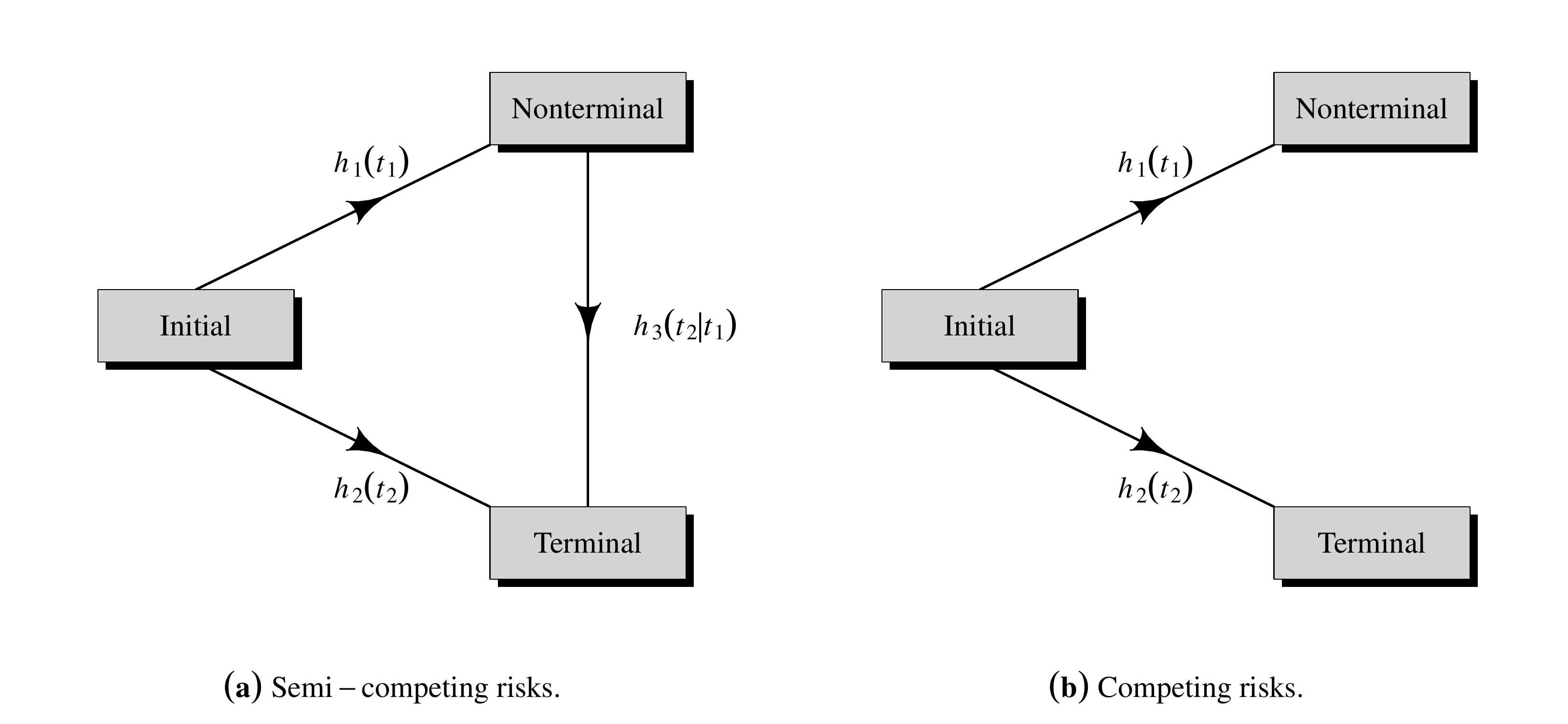}
	\caption{Graphical representation of \textbf{(a)} semi-competing risks and \textbf{(b)} competing risks.}
	\label{fig:Figure1}
\end{figure}

Toward formally describing the structure of semi-competing risks data, let $T_{1}$ and $T_{2}$ denote the times to the non-terminal and terminal events, respectively. From the modeling perspective, the focus in the semi-competing risks setting is to characterize the distribution $T_{1}$ and its potential relationship with the distribution of $T_{2}$, i.e. the joint distribution of ($T_1$, $T_2$). For example, from an initial state (e.g., transplantation), as time progresses, a subject could make a transition into the non-terminal or terminal state (see Figure~\ref{fig:Figure1}.a). In the case of a transition into the non-terminal state, the subject could subsequently transition into the terminal state even if these transitions cannot occur in the reverse order. The main disadvantage of the competing risks framework (see Figure~\ref{fig:Figure1}.b) to the study of non-terminal event is that it does not utilize the information on the occurrence and timing of terminal event following the non-terminal event, which could be used to understand the dependence between the two events.

The current literature for the analysis of semi-competing risks data is composed of three approaches: methods that specify the dependence between non-terminal and terminal events via a copula \citep{fine2001, wang2003, jiang2005, ghosh2006, peng2007, lakhal2008, hsieh2008, fu2013}; methods based on multi-state models, specifically the so-called \textit{illness-death} model \citep{liu2004, putter2007, ye2007, kneib2008, zeng2009, xu2010, zeng2012, han2014, zhang2014, khlee2015, khlee2016}; and methods built upon the principles of causal inference \citep{zhang2003, egleston2007, tchetgen2014, varadhan2014}.  

The \textbf{SemiCompRisks} package is designed to provide a comprehensive suite of functions for the analysis of semi-competing risks data based on the illness-death model, together with, as a complementary suite of tools, functions for the analysis of univariate time-to-event data. While Bayesian methods are used for estimation and inference for all available models, maximum likelihood estimation is also provided for select parametric models. Furthermore, \textbf{SemiCompRisks} offers flexible parametric and non-parametric specifications for baseline survival functions and cluster-specific random effects distributions under accelerated failure time and proportional hazards models. The functionality of the package covers methods proposed in a series of recent papers on the analysis of semi-competing risks data~\citep{khlee2015, khlee2016, khlee2017a}.

The remainder of the paper is organized as follows. Section~\ref{sec:packages} summarizes existing R packages that provide methods for multi-state modeling, and explains the key contributions of the \textbf{SemiCompRisks} package. Section~\ref{sec:data} introduces an on-going study of stem cell transplantation and provides a description of the data available in the package. Section~\ref{sec:model} presents different specifications of models and estimation methods implemented in our package. Section~\ref{sec:descr} summarizes the core components of the \textbf{SemiCompRisks} package, including datasets, functions for fitting models, functions, the structure of output provided to analysts. Section~\ref{sec:applic} illustrates the usage of the main functions in the package through three semi-competing risks analyses of the stem cell transplantation data. Finally, Section~\ref{sec:disc} concludes with discussion and an overview of the extensions we are working on.

\section{Other packages and their features} \label{sec:packages}

As we elaborate upon below, the illness-death model for semi-competing risks, that is the focus on the \textbf{SemiCompRisks} package, is a special case of the broader class of multi-state models. Currently, there are numerous R packages that permit estimation and inference for a multi-state model and that could conceivably be used to analyze semi-competing risks data.

The \textbf{mvna} package computes the Nelson-Aalen estimator of the cumulative transition hazard for arbitrary Markov multi-state models with right-censored and left-truncated data, but it does not compute transition probability matrices \citep{allignol2008}. The \textbf{TPmsm} implements non-parametric and semi-parametric estimators for the transition probabilities in 3-state models, including the Aalen-Johansen estimator and estimators that are consistent even without Markov assumption or in case of dependent censoring \citep{araujo2014}. The \textbf{p3state.msm} package performs inference in an illness-death model \citep{meira2011}. Its main feature is the ability for obtaining non-Markov estimates for the transition probabilities. The \textbf{etm} package calculates the empirical transition probability matrices and corresponding variance estimates for any time-inhomogeneous multi-state model with finite state space and data subject to right-censoring and left-truncation, but it does not account for the influence of covariates \citep{allignol2011}. The \textbf{msm} package is able to fit time-homogeneous Markov models to panel count data and hidden Markov models in continuous time \citep{jackson2011}. The time-homogeneous Markov approach could be a particular case of the illness-death model, where interval-censored data can be considered. The \textbf{tdc.msm} package may be used to fit the time-dependent proportional hazards model and multi-state regression models in continuous time, such as Cox Markov model, Cox semi-Markov model, homogeneous Markov model, non-homogeneous piecewise model, and non-parametric Markov model \citep{meira2007}. The \textbf{SemiMarkov} package performs parametric (Weibull or exponentiated Weibull specification) estimation in a homogeneous semi-Markov model \citep{krol2015}. Moreover, the effects of covariates on the process evolution can be studied using a semi-parametric Cox model for the distributions of sojourn times. The \textbf{flexsurv} package provides functions for fitting and predicting from fully-parametric multi-state models with Markov or semi-Markov specification \citep{jackson2016}. In addition, the multi-state models implemented in \textbf{flexsurv} give the possibility to include interval-censoring and some of them also left-truncation. The \textbf{msSurv} calculates non-parametric estimation of general multi-state models subject to independent right-censoring and possibly left-truncation \citep{ferguson2012}. This package also computes the marginal state occupation probabilities along with the corresponding variance estimates, and lower and upper confidence intervals. The \textbf{mstate} package can be applied to right-censored and left-truncated data in semi-parametric or non-parametric multi-state models with or without covariates and it may also be used to competing risk models \citep{wreede2011}. Specifically for Cox-type illness-death models to interval-censored data, we highlight the packages \textbf{coxinterval} \citep{boruvka2015} and \textbf{SmoothHazard} \citep{touraine2017}, where the latter also allows that the event times to be left-truncated. Finally, \textbf{frailtypack} package permits the analysis of correlated data under select clusterings, as well as the analysis of left-truncated data, through a focus on frailty models using penalized likelihood estimation or parametric estimation \citep{rondeau2012}.

While these packages collectively provide broad functionality, each of them is either non-specific to semi-competing risks or only permits consideration of a narrow model specifications. In developing the \textbf{SemiCompRisks} package, the goal was to provide a single package within which a broad range of models and model specifications could be entertained. The \textbf{frailtypack} package, for example, can also be used to analyze cluster-correlated semi-competing risks data but it is restricted to the proportional hazards model with either patient-specific or cluster-specific random effects but not both~\citep{liquet2012}. Furthermore, estimation/inference is within the frequentist framework so that estimation of hospital-specific random effects, of particular interest in health policy applications~\citep{khlee2016}, together with the quantification of uncertainty is incredibly challenging. This, however, is (relatively) easily achieved through the functionality of \textbf{SemiCompRisks} package. Given the breadth of the functionality of the package, in addition to the usual help files, we have developed a series of model-specific vignettes which can be accessed through the CRAN \citep{khlee2017b} or R command \texttt{vignette("SemiCompRisks")}, covering a total of 12 distinct model specifications.

\section{CIBMTR data} \label{sec:data}

The example dataset used throughout this paper was obtained from the Center for International Blood and Marrow Transplant Research (CIBMTR), a collaboration between the National Marrow Donor Program and the Medical College of Wisconsin representing a worldwide network of transplant centers \citep{clee2017}. For illustrative purposes, we consider a hypothetical study in which the goal is to investigate risk factors for grade III or IV acute graft-versus-host disease (GVHD) among $9,651$ patients who underwent the first allogeneic hematopoietic cell transplant (HCT) between January $1999$ and December $2011$. 

As summarized in Table~\ref{tab:description}, after administratively censoring follow-up at $365$ days post-transplant, each patient can be categorized according to their observed outcome information into four groups: (i) acute GVHD and death; (ii) acute GVHD and censored for death; (iii) death without acute GVHD; and (iv) censored for both. Furthermore, for each patient, the following covariates are available:gender (Male, Female); age ($<$10, 10-19, 20-29, 30-39, 40-49, 50-59, 60+); disease type (AML, ALL, CML, MDS); disease stage (Early, Intermediate, Advanced); and HLA compatibility (Identical sibling, 8/8, 7/8).

We note that due to confidentiality considerations the original study outcomes (\texttt{time1}, \texttt{time2}, \texttt{event1}, \texttt{event2}: times and censoring indicators to the non-terminal and terminal events) are not available in \textbf{SemiCompRisks} package. As such we provide the five original covariates together with estimates of parameters from the analysis of CIBMTR data, so that one could simulate semi-competing risks outcomes (see the simulation procedure in Appendix~\ref{app:sec2}). Based on this, the data shown in Table~\ref{tab:description} reflects simulated outcome data using $1405$ as the seed.

\begin{table}[htb]
\centering \small
\begin{tabular}{lrccccc}
\bottomrule
& & & \multicolumn{4}{c}{Outcome category (\%)}\\
\cline{4-7}
 &   	 &  	& \multirow{3}{*}{\shortstack{Both \\ acute GVHD \\ \& death}} & \multirow{3}{*}{\shortstack{Acute GVHD \\ \& censored \\ for death}} & \multirow{3}{*}{\shortstack{Death \\ without \\ acute GVHD}} & \multirow{3}{*}{\shortstack{Censored \\ for both}} \\
 & \multirow{2}{*}{\shortstack{$N$}~~} & \multirow{2}{*}{\shortstack{\%}} &     &   &  	 &  \\
 &  &  & 	&  	&  &  \\
\toprule
Total subjects 		& 	9,651 & 100.0 & ~9.5 & ~8.9 & 28.8 & 52.8 \\
\texttt{Gender} 		&  &  &  &  &  &  \\
~~~Male   		& 5,366 & ~55.6 	& ~9.7 & ~9.5 & 28.1 & 52.7 \\
~~~Female 		& 4,285 & ~44.4 	& ~9.1 & ~8.3 & 29.7 & 52.9 \\
\texttt{Age}, years 	&  &  &  &  &  &  \\
~~~$<$10 		  &  653  & ~~6.8 	&  ~5.0 & 11.9 & 23.4 & 59.7 \\
~~~10-19 			& 1,162 & ~12.0 	&  ~8.0 & 11.4 & 24.0 & 56.6 \\
~~~20-29 			& 1,572 & ~16.3 	&  ~9.7 & ~9.9 & 27.4 & 53.0 \\
~~~30-39 			& 1,581 & ~16.4 	&  ~9.8 & 10.7 & 28.5 & 51.0 \\
~~~40-49 			& 2,095 & ~21.7 	&  11.0 & ~9.6 & 29.7 & 49.7 \\
~~~50-59 			& 2,008 & ~20.8 	&  ~9.8 & ~5.1 & 32.3 & 52.8 \\
~~~60+   			&  580  & ~~6.0 	&  ~9.9 & ~4.8 & 33.1 & 52.2 \\
\texttt{Disease type} &  &  &  &  &  &  \\
~~~AML 			& 4,919 & ~51.0  	& ~8.2 & ~8.0 & 30.3 & 53.5 \\
~~~ALL 			& 2,071 & ~21.5  	& ~9.9 & ~9.0 & 29.3 & 51.8 \\
~~~CML 			& 1,525 & ~15.8  	& 12.1 & 11.3 & 22.2 & 54.4 \\
~~~MDS 			& 1,136 & ~11.8  	& 11.0 & 10.0 & 30.0 & 49.0 \\	
\texttt{Disease status} &  &  &  &  &  &  \\
~~~Early        & 4,873 & ~50.5 	& ~8.4 & 11.0 & 23.6 & 57.0 \\
~~~Intermediate & 2,316 & ~24.0 	& ~9.7 & ~8.5 & 30.1 & 51.7 \\
~~~Advanced     &	2,462 & ~25.5 	& 11.5 & ~5.4 & 37.7 & 45.4 \\
\texttt{HLA compatibility} &  &  &  &  &  &  \\
~~~Identical sibling 	& 3,941 & ~40.8 	&  ~7.4 & ~8.5 & 26.3 & 57.8 \\
~~~8/8 			& 4,100 & ~42.5 	& 10.5 & ~9.7 & 30.3 & 49.5 \\
~~~7/8 			& 1,610 & ~16.7 	& 12.2 & ~8.1 & 30.9 & 48.8 \\
\bottomrule
\end{tabular}
\caption{Covariate and simulated outcome information for 9,651 patients who underwent the first HCT between 1999-2011 with administrative censoring at 365 days.}
\label{tab:description}
\end{table}

\section{The illness-death models for semi-competing risks data} \label{sec:model}

We offer three flexible multi-state illness-death models for the analysis of semi-competing risks data: accelerated failure time (AFT) models for independent data; proportional hazards regression (PHR) models for independent data; and PHR models for cluster-correlated data. These models accommodate parametric or non-parametric specifications for baseline survival functions as well as a Markov or semi-Markov assumptions for terminal event following non-terminal event.

\subsection{AFT models for independent semi-competing risks data} \label{sec:AFTind}

In the AFT model specification, we directly model the connection between event times and covariates \citep{wei1992}. For the analysis of semi-competing risks data, we consider the following AFT model specifications under the illness-death modeling framework \citep{khlee2017a}:
\begin{eqnarray}
	\log(T_{i1}) 				&=& \bm{x}_{i1}^{\top}\bm{\beta}_{1} + \gamma_{i} + \epsilon_{i1}, ~~ T_{i1} > 0, 		 \label{eq:AFTind1} \\
	\log(T_{i2}) 				&=& \bm{x}_{i2}^{\top}\bm{\beta}_{2} + \gamma_{i} + \epsilon_{i2}, ~~ T_{i2} > 0, 		 \label{eq:AFTind2} \\
	\log(T_{i2}-T_{i1}) &=& \bm{x}_{i3}^{\top}\bm{\beta}_{3} + \gamma_{i} + \epsilon_{i3}, ~~ T_{i2} > T_{i1}, \label{eq:AFTind3}
\end{eqnarray}

\noindent where $T_{i1}$ and $T_{i2}$ denote the times to the non-terminal and terminal events, respectively, from subject $i=1,\ldots,n$, $\bm{x}_{ig}$ is a vector of transition-specific covariates, $\bm{\beta}_{g}$ is a corresponding vector of transition-specific regression parameters, and $\epsilon_{ig}$ is a transition-specific random variable whose distribution determines that of the corresponding transition time, $g \in \left\{1, 2, 3\right\}$. Finally, in each of (\ref{eq:AFTind1})-(\ref{eq:AFTind3}), $\gamma_{i}$ is a study subject-specific random effect that induces positive dependence between the two event times. We assume that $\gamma_{i}$ follows a Normal(0, $\theta$) distribution and adopt a conjugate inverse Gamma distribution, denoted by IG$(a^{(\theta)}, b^{(\theta)})$ for the variance component $\theta$. For regression parameters $\bm{\beta}_{g}$, we adopt non-informative flat prior on the real line.

From models (\ref{eq:AFTind1})-(\ref{eq:AFTind3}), we can adopt either a fully parametric or a semi-parametric approach depending on the specification of the distributions for $\epsilon_{i1}$, $\epsilon_{i2}$, $\epsilon_{i3}$. We build a parametric modeling based on the log-Normal formulation, where $\epsilon_{ig}$ follows a Normal$(\mu_{g}, \sigma^{2}_{g})$ distribution. We adopt non-informative flat priors on the real line for $\mu_{g}$ and independent IG$(a_{g}^{(\sigma)},b_{g}^{(\sigma)})$ for $\sigma_{g}^{2}$. As an alternative, a semi-parametric framework can be considered by adopting independent non-parametric Dirichlet process mixtures (DPM) of $M_{g}$ Normal$(\mu_{gr}, \sigma^{2}_{gr})$ distributions, $r \in \left\{1,\ldots,M_{g}\right\}$, for each $\epsilon_{ig}$. Following convention in the literature, we refer to each component Normal distribution as being specific to some ``class'' \citep{neal2000}. Since the class-specific $(\mu_{gr}, \sigma^{2}_{gr})$ are unknown, they are assumed to be draws from a so-called the \emph{centering distribution}. Specifically, we take a Normal distribution centered at $\mu_{g0}$ with a variance $\sigma_{g0}^{2}$ for $\mu_{gr}$ and an IG$(a_{g}^{(\sigma_{gr})}, b_{g}^{(\sigma_{gr})})$ for $\sigma_{gr}^{2}$. Furthermore, since the ``true'' class membership for any given study subject is unknown, we let $p_{gr}$ denote the probability of belonging to the $r$th class for transition $g$ and $\bm{p}_{g}=(p_{g1},\ldots,p_{gM_{g}})^{\top}$ the collection of such probabilities. In the absence of prior knowledge regarding the distribution of class memberships for the $n$ subjects across the $M_{g}$ classes, $\bm{p}_{g}$ is assumed to follow a conjugate symmetric Dirichlet$(\tau_{g}/M_{g},\ldots,\tau_{g}/M_{g})$ distribution, where $\tau_{g}$ is referred to as the \emph{precision parameter} \citep[for more details, see][]{khlee2017a}.

Our AFT modeling framework can also handle interval-censored and/or left-truncated semi-competing risks data. Suppose that subject $i$ was observed at follow-up times $\left\{c_{i1},\ldots, c_{im_{i}}\right\}$ and let $c_{i}^{*}$ and $L_{i}$ denote the time to the end of study (or administrative right-censoring) and the time at study entry (i.e., the left-truncation time), respectively. Considering interval-censoring for both events, $T_{i1}$ and $T_{i2}$, for $i=1,\ldots,n$, satisfy $c_{ij} \leq T_{i1} < c_{ij+1}$ for some $j$ and $c_{ik} \leq T_{i2} < c_{ik+1}$ for some $k$, respectively. Therefore, the observed outcome information for interval-censored and left-truncated semi-competing risks data for the subject $i$ can be represented by $\{L_i, c_{ij}, c_{ij+1}, c_{ik}, c_{ik+1}\}$.

\subsection{PHR models for independent semi-competing risks data} \label{sec:PHRind}

We consider an illness-death multi-state model with proportional hazards assumptions characterized by three hazard functions (see Figure~\ref{fig:Figure1}.a) that govern the rates at which subjects transition between the states: a cause-specific hazard for non-terminal event, $h_{1}(t_{i1})$; a cause-specific hazard for terminal event, $h_{2}(t_{i2})$; and a hazard for terminal event conditional on a time for non-terminal event, $h_{3}(t_{i2} \mid t_{i1})$. We consider the following specification for hazard functions \citep{xu2010, khlee2015}:
\begin{eqnarray}
	h_{1}(t_{i1} \mid \gamma_{i}, \bm{x}_{i1}) 				 &=& \gamma_{i}\,h_{01}(t_{i1})\exp(\bm{x}_{i1}^{\top}\bm{\beta}_{1}), ~~ t_{i1} > 0, 		  				 \label{eq:PHRind1} \\
	h_{2}(t_{i2} \mid \gamma_{i}, \bm{x}_{i2}) 				 &=& \gamma_{i}\,h_{02}(t_{i2})\exp(\bm{x}_{i2}^{\top}\bm{\beta}_{2}), ~~ t_{i2} > 0, 		  				 \label{eq:PHRind2} \\
	h_{3}(t_{i2} \mid t_{i1}, \gamma_{i}, \bm{x}_{i3}) &=& \gamma_{i}\,h_{03}(z(t_{i1},t_{i2}))\exp(\bm{x}_{i3}^{\top}\bm{\beta}_{3}), ~~ t_{i2} > t_{i1}, \label{eq:PHRind3}
\end{eqnarray}

\noindent where $h_{0g}$ is an unspecified baseline hazard function and $\bm{\beta}_{g}$ is a vector of log-hazard ratio regression parameters associated with the covariates $\bm{x}_{ig}$. Finally, in each of (\ref{eq:PHRind1})-(\ref{eq:PHRind3}), $\gamma_{i}$ is a study subject-specific shared frailty following a Gamma($\theta^{-1}$, $\theta^{-1}$) distribution, parametrized so that $E\left[\gamma_{i}\right]=1$ and $V\left[\gamma_{i}\right]=\theta$. The model (\ref{eq:PHRind3}) is referred to as being Markov or semi-Markov depending on whether we assume $z(t_{i1},t_{i2})=t_{i2}$ or $z(t_{i1},t_{i2})=t_{i2}-t_{i1}$, respectively.

The Bayesian approach for models (\ref{eq:PHRind1})-(\ref{eq:PHRind3}) requires the specification of prior distributions for unknown parameters. For the regression parameters $\bm{\beta}_{g}$, we adopt a non-informative flat prior distribution on the real line. For the variance in the subject-specific frailties, $\theta$, we adopt a Gamma$(a^{(\theta)}, b^{(\theta)})$ for the precision $\theta^{-1}$. For the parametric specification for baseline hazard functions, we consider a Weibull model: $h_{0g}(t)=\alpha_{g} \, \kappa_{g} \, t^{\alpha_{g}-1}$. We assign a Gamma$(a_{g}^{(\alpha)},b_{g}^{(\alpha)})$ for $\alpha_{g}$ and a Gamma$(c_{g}^{(\kappa)},d_{g}^{(\kappa)})$ for $\kappa_{g}$. As an alternative, a non-parametric piecewise exponential model (PEM) is considered for baseline hazard functions based on taking each of the log-baseline hazard functions to be a flexible mixture of piecewise constant function. Let $s_{g,\mbox{max}}$ denote the largest observed event time for each transition and construct a finite partition of the time axis, $0=s_{g,0} < s_{g,1} < s_{g,2} < \ldots < s_{g,K_{g}+1} = s_{g,\mbox{max}}$. Letting $\bm{\lambda}_{g}=(\lambda_{g,1},\ldots,\lambda_{g,K_{g}},\lambda_{g,K_{g}+1})^{\top}$ denote the heights of the log-baseline hazard function on the disjoint intervals based on the time splits $\bm{s}_{g}=(s_{g,1}, \ldots, s_{g,K_{g}+1})^{\top}$, we assume that $\bm{\lambda}_{g}$ follows a multivariate Normal distribution (MVN), MVN$(\mu_{\lambda_{g}}\bm{1},\sigma_{\lambda_{g}}^{2}\bm{\Sigma}_{\lambda_{g}})$, where $\mu_{\lambda_{g}}$ is the overall mean, $\sigma_{\lambda_{g}}^{2}$ represents a common variance component for the $K_{g}+1$ elements, and $\bm{\Sigma}_{\lambda_{g}}$ specifies the covariance structure these elements. We adopt a flat prior on the real line for $\mu_{\lambda_{g}}$ and a conjugate Gamma$(a_{g}^{(\sigma)},b_{g}^{(\sigma)})$ distribution for the precision $\sigma_{\lambda_{g}}^{-2}$. In order to relax the assumption of fixed partition of the time scales, we adopt a Poisson$(\alpha_{g}^{(K)})$ prior for the number of splits, $K_{g}$, and conditioned on the number of splits, we consider locations, $\bm{s}_{g}$, to be \emph{a priori} distributed as the even-numbered order statistics:
\begin{equation}
\pi(\bm{s}_{g} \mid K_{g}) \propto \frac{(2K_{g}+1)!\prod_{k=1}^{K_{g}+1}(s_{g,k}-s_{g,k-1})}{(s_{g,K_{g}+1})^{2K_{g}+1}}. \label{eq:locpartition}
\end{equation} 
Note that the prior distributions of $K_{g}$ and $\bm{s}_{g}$ jointly form a time-homogeneous Poisson process prior for the partition $(K_{g},\bm{s}_{g})$. For more details, see \cite{khlee2015}.

\subsection{PHR models for cluster-correlated semi-competing risks data} \label{sec:PHRcluster}

\cite{khlee2016} proposed hierarchical models that accommodate correlation in the joint distribution of the non-terminal and terminal events across patients for the setting where patients are clustered within hospitals. The hierarchical models for cluster-correlated semi-competing risks data build upon the illness-death model given in (\ref{eq:PHRind1})-(\ref{eq:PHRind3}). Let $T_{ji1}$ and $T_{ji2}$ denote the times to the non-terminal and terminal events for the $i$th subject in the $j$th cluster, respectively, for $i=1,\ldots,n_{j}$ and $j=1,\ldots,J$. The general modeling specification is given by:
\begin{eqnarray}
	h_{1}(t_{ji1} \mid \gamma_{ji}, \bm{x}_{ji1}, V_{j1}) 				 &=& \gamma_{ji}\,h_{01}(t_{ji2})\exp(\bm{x}_{ji1}^{\top}\bm{\beta}_{1} + V_{j1}), ~~ t_{ji1} > 0, 		  						 \label{eq:PHRclus1} \\
	h_{2}(t_{ji2} \mid \gamma_{ji}, \bm{x}_{ji2}, V_{j2}) 				 &=& \gamma_{ji}\,h_{02}(t_{ji2})\exp(\bm{x}_{ji2}^{\top}\bm{\beta}_{2} + V_{j2}), ~~ t_{ji2} > 0, 		  						 \label{eq:PHRclus2} \\
	h_{3}(t_{ji2} \mid t_{ji1}, \gamma_{ji}, \bm{x}_{ji3}, V_{j3}) &=& \gamma_{ji}\,h_{03}(z(t_{ji1},t_{ji2}))\exp(\bm{x}_{ji3}^{\top}\bm{\beta}_{3} + V_{j3}), ~~ t_{ji2} > t_{ji1}, \label{eq:PHRclus3}
\end{eqnarray}

\noindent where $h_{0g}$ is an unspecified baseline hazard function and $\bm{\beta}_{g}$ is a vector of log-hazard ratio regression parameters associated with the covariates $\bm{x}_{jig}$. A study subject-specific shared frailty $\gamma_{ji}$ is assumed to follow a Gamma($\theta^{-1}$, $\theta^{-1}$) distribution and $\bm{V}_{j}=(V_{j1}, V_{j2}, V_{j3})^{\top}$ is a vector of cluster-specific random effects, each specific to one of the three possible transitions.

From a Bayesian perspective for models (\ref{eq:PHRclus1})-(\ref{eq:PHRclus3}), we can adopt either a parametric Weibull or non-parametric PEM specification for baseline hazard functions $h_{0g}$ with their respective configurations of prior distributions analogous to those outlined in Section \ref{sec:PHRind}. For the parametric specification of cluster-specific random effects, we assume that $\bm{V}_{j}$ follows MVN$_{3}(\bm{0},\bm{\Sigma}_{V})$ distribution. We adopt a conjugate inverse-Wishart$(\bm{\Psi}_{v},\rho_{v})$ prior for the variance-covariance matrix $\bm{\Sigma}_{V}$. For the non-parametric specification, we adopt a DPM of MVN distributions with a centering distribution, $G_{0}$, and a precision parameter, $\tau$. Here we take $G_{0}$ to be a multivariate Normal/inverse-Wishart (NIW) distribution for which the probability density function can be expressed as the product:
\begin{equation}
f_{\mbox{NIW}}(\bm{\mu},\bm{\Sigma} \mid \bm{\Psi}_{0},\rho_{0}) = f_{\mbox{MVN}}(\bm{\mu} \mid \bm{0}, \bm{\Sigma}) \times f_{\mbox{inverse-Wishart}}(\bm{\Sigma} \mid \bm{\Psi}_{0}, \rho_{0}), \label{eq:NIW}
\end{equation}
where $\bm{\Psi}_{0}$ and $\rho_{0}$ are the hyperparameters of $f_{\mbox{NIW}}(\cdot)$. We assign a Gamma$(a_{\tau},b_{\tau})$ prior distribution for $\tau$. Finally, for $\bm{\beta}_{g}$ and $\theta$, we adopt the same priors as those adopted for the model in Section \ref{sec:PHRind}. For more details, see \cite{khlee2016}.

\subsection{Estimation and inference}

Bayesian estimation and inference is available for all models in the \textbf{SemiCompRisks}. Additionally, one may also choose to use maximum likelihood estimation for the parametric Weibull PHR model described in Section \ref{sec:PHRind}.

To perform Bayesian estimation and inference, we use a random scan Gibbs sampling algorithm to generate samples from the full posterior distribution. Depending on the complexity of the model adopted, the Markov chain Monte Carlo (MCMC) scheme may also include additional strategies, such as Metropolis-Hastings and reversible jump MCMC (Metropolis-Hastings-Green) steps. Specific details of each implementation can be seen in the online supplemental materials of \cite{khlee2015, khlee2016, khlee2017a}.

\section{Package description} \label{sec:descr}

The \textbf{SemiCompRisks} package contains three key functions, \texttt{FreqID\_HReg}, \texttt{BayesID\_HReg} and \texttt{BayesID\_AFT}, focused on models for semi-competing risks data as well as the analogous univariate survival models, \texttt{FreqSurv\_HReg}, \texttt{BayesSurv\_HReg} and \texttt{BayesSurv\_AFT}. It also provides two auxiliary functions, \texttt{initiate.startValues\_HReg} and \texttt{initiate.startValues\_AFT}, that can be used to generate initial values for Bayesian estimation; \texttt{simID} and \texttt{simSurv} functions for simulating semi-competing risks and univariate survival data, respectively; five covariates and parameter estimates from \texttt{CIBMTR} data; and the \texttt{BMT} dataset referring to 137 bone marrow transplant patients.

\subsection{Summary of functionality} \label{subsec:func}

Table~\ref{tab:modelsimplemented} shows the modeling options implemented in the \textbf{SemiCompRisks} package for both semi-competing risks and univariate analysis. Specifically, we categorize the approaches based on the analysis type (semi-competing risks or univariate), the survival model (AFT or PHR), data type (independent or clustered), accommodation to left-truncation and/or interval-censoring in addition to right-censoring, and also statistical paradigms (frequentist or Bayesian).

\begin{table}[htb]
\centering
\begin{threeparttable}
\begin{tabular}{|c|c|c|c|c|}
\hline
Analysis & Model & Data type   & L-T and/or I-C   & Statistical paradigm \\
\hline	
								& \multirow{4}{*}{AFT}&  \multirow{2}{*}{Independent} &  No	& B \\ \cline{4-5}
								& & 					&   Yes	  & B \\ \cline{3-5}
								& & \multirow{2}{*}{Clustered}	&    No		 & x \\ \cline{4-5}
Semi-competing					& & 					&   Yes   & x \\ \cline{2-5}
risks								& \multirow{4}{*}{PHR} & \multirow{2}{*}{Independent} & No & B \& F \\ \cline{4-5}
								& & 				  &   Yes 	& x \\ \cline{3-5}
								& & \multirow{2}{*}{Clustered} & No & B \\ \cline{4-5}
								& & 				  &	  Yes 	& x \\
\hline								
\multirow{8}{*}{Univariate}	& \multirow{4}{*}{AFT}&  \multirow{2}{*}{Independent} &  No	& B \\ \cline{4-5}
								& & 					&   Yes	  & B \\ \cline{3-5}
								& & \multirow{2}{*}{Clustered}	&    No		 & x \\ \cline{4-5}
								& & 					&   Yes   & x \\ \cline{2-5}
								& \multirow{4}{*}{PHR} & \multirow{2}{*}{Independent} & No & B \& F \\ \cline{4-5}
								& & 				  &   Yes 	& x \\ \cline{3-5}
								& & \multirow{2}{*}{Clustered} & No & B \\ \cline{4-5}
								& & 				  &	  Yes 	& x \\								
\hline			
\end{tabular}\begin{tablenotes}
      \footnotesize
      \item L-T: left-truncation; \hspace{0.1cm} I-C: interval-censoring; \hspace{0.1cm} B: Bayesian; \hspace{0.1cm} F: frequentist; \hspace{0.1cm} x: not available
\end{tablenotes}
\end{threeparttable}
\caption{Models implemented in the \textbf{SemiCompRisks} package.}
\label{tab:modelsimplemented}
\end{table}

The full description of functionality of the \textbf{SemiCompRisks} package can be accessed through the R command \texttt{help("SemiCompRisks")} or \texttt{vignette("SemiCompRisks")} which provides in detail the specification of all models implemented in the package. Below we describe the input data format and some crucial arguments for defining and fitting a model for semi-competing risks data using the \textbf{SemiCompRisks} package.

\subsubsection{Model specification} \label{subsec:data}

From a semi-competing risks dataset, we jointly define the outcomes and covariates in a \texttt{Formula} object. Here we use the \texttt{simCIBMTR} dataset, obtained from the simulation procedure presented in Appendix~\ref{app:sec2}:
\begin{verbatim}
R> form <- Formula(time1 + event1 | time2 + event2 ~ dTypeALL + dTypeCML + 
+     dTypeMDS + sexP | dTypeALL + dTypeCML + dTypeMDS | dTypeALL +
+     dTypeCML + dTypeMDS)
\end{verbatim}

The outcomes \texttt{time1}, \texttt{time2}, \texttt{event1} and \texttt{event2} denote the times and censoring indicators to the non-terminal and terminal events, respectively, and the covariates of each hazard function are separated by $|$ (vertical bar).

The specification of the \texttt{Formula} object varies slightly if the semi-competing risks model accommodates left-truncated and/or interval-censored data (see vignette documentation \cite{khlee2017b}).

\subsubsection{Critical arguments} \label{subsec:critical}

Most functions for semi-competing risks analysis in the \textbf{SemiCompRisks} package take common arguments. These arguments and their descriptions are shown as follows:

\begin{itemize}
	\item \texttt{id}: a vector of cluster information for $n$ subjects, where cluster membership corresponds to one of the positive integers $1,\ldots,J$. 
	\item \texttt{model}: a character vector that specifies the type of components in a model. It can have up to three elements depending on the model specification. The first element is for the assumption on $h_{3}$: ``semi-Markov'' or ``Markov''. The second element is for the specification of baseline hazard functions for PHR models - ``Weibull'' or ``PEM'' - or baseline survival distribution for AFT models - ``LN'' (log-Normal) or ``DPM''. The third element needs to be set only for clustered semi-competing risks data and is for the specification of cluster-specific random effects distribution: ``MVN'' or ``DPM''.
	\item \texttt{hyperParams}: a list containing vectors for hyperparameter values in hierarchical models.
	\item \texttt{startValues}: a list containing vectors of starting values for model parameters.
	\item \texttt{mcmcParams}: a list containing variables required for MCMC sampling.
\end{itemize}

Hyperparameter values, starting values for model parameters, and MCMC arguments depend on the specified Bayesian model and the assigned prior distributions. For a list of illustrations, see vignette documentation \cite{khlee2017b}.

\subsection[FreqIDHReg]{\texttt{FreqID\_HReg}} \label{sec:FHRegfc}

The function \texttt{FreqID\_HReg} fits Weibull PHR models for independent semi-competing risks data, as in (\ref{eq:PHRind1})-(\ref{eq:PHRind3}), based on maximum likelihood estimation. Its default structure is given by
\begin{verbatim}
FreqID_HReg(Formula, data, model="semi-Markov", frailty=TRUE),
\end{verbatim}
\noindent where \texttt{Formula} represents the outcomes and the linear predictors jointly, as presented in Section~\ref{subsec:func}; \texttt{data} is a data frame containing the variables named in \texttt{Formula}; \texttt{model} is one of the critical arguments of the \textbf{SemiCompRisks} package (see Section~\ref{subsec:func}), in which it specifies the type of model based on the assumption on $h_{3}(t_{i2} \mid t_{i1}, \cdot)$ in (\ref{eq:PHRind3}). Here, \texttt{model} can be ``\texttt{Markov}'' or ``\texttt{semi-Markov}''. Finally, \texttt{frailty} is a logical value (\texttt{TRUE} or \texttt{FALSE}) to determine whether to include the subject-specific shared frailty term $\gamma$ into the illness-death model.

\subsection[BayesIDHReg]{\texttt{BayesID\_HReg}} \label{sec:BHRegfc}

The function \texttt{BayesID\_HReg} fits parametric and semi-parametric PHR models for independent or cluster-correlated semi-competing risks data, as in (\ref{eq:PHRind1})-(\ref{eq:PHRind3}) or (\ref{eq:PHRclus1})-(\ref{eq:PHRclus3}), based on Bayesian inference. Its default structure is given by
\begin{verbatim}
BayesID_HReg(Formula, data, id=NULL, model=c("semi-Markov","Weibull"),
hyperParams, startValues, mcmcParams, path=NULL)
\end{verbatim}

\texttt{Formula} and \texttt{data} are analogous to the previous case; \texttt{id}, \texttt{model}, \texttt{hyperParams}, \texttt{startValues}, and \texttt{mcmcParams} are all critical arguments of the \textbf{SemiCompRisks} package (see Section \ref{subsec:func}), where \texttt{id} indicates the cluster that each subject belongs to (for independent data, \texttt{id=NULL}); \texttt{model} allows us to specify either ``\texttt{Markov}'' or ``\texttt{semi-Markov}'' assumption, whether the priors for baseline hazard functions are parametric (``\texttt{Weibull}'') or non-parametric (``\texttt{PEM}''), and whether the cluster-specific random effects distribution is parametric (``\texttt{MVN}'') or non-parametric (``\texttt{DPM}''). The third element of \texttt{model} is only required for models for clustered-correlated data given in (\ref{eq:PHRclus1})-(\ref{eq:PHRclus3}).

The \texttt{hyperParams} argument defines all model hyperparameters: \texttt{theta} (a numeric vector for hyperparameters, $a^{(\theta)}$ and $b^{(\theta)}$, in the prior of subject-specific frailty variance component), \texttt{WB} (a list containing numeric vectors for Weibull hyperparameters ($a_{g}^{(\alpha)}$, $b_{g}^{(\alpha)}$) and ($c_{g}^{(\kappa)}$, $d_{g}^{(\kappa)}$) for $g \in \left\{1, 2, 3\right\}$: \texttt{WB.ab1}, \texttt{WB.ab2}, \texttt{WB.ab3}, \texttt{WB.cd1}, \texttt{WB.cd2}, \texttt{WB.cd3}), \texttt{PEM} (a list containing numeric vectors for PEM hyperparameters ($a_{g}^{(\sigma)}$, $b_{g}^{(\sigma)}$), and $\alpha_{g}^{(K)}$ for $g \in \left\{1, 2, 3\right\}$: \texttt{PEM.ab1}, \texttt{PEM.ab2}, \texttt{PEM.ab3}, \texttt{PEM.alpha1}, \texttt{PEM.alpha2}, \texttt{PEM.alpha3}); and for the analysis of clustered semi-competing risks data, additional components are required: \texttt{MVN} (a list containing numeric vectors for MVN hyperparameters $\bm{\Psi}_{v}$ and $\rho_{v}$: \texttt{Psi\_v}, \texttt{rho\_v}), \texttt{DPM} (a list containing numeric vectors for DPM hyperparameters $\bm{\Psi}_{0}$, $\rho_{0}$, $a_{\tau}$, and $b_{\tau}$: \texttt{Psi0}, \texttt{rho0}, \texttt{aTau}, \texttt{bTau}).

The \texttt{startValues} argument specifies initial values for model parameters. This specification can be done manually or through the auxiliary function \texttt{initiate.startValues\_HReg}. The \texttt{mcmcParams} argument sets the information for MCMC sampling: \texttt{run} (a list containing numeric values for setting for the overall run: \texttt{numReps}, total number of scans; \texttt{thin}, extent of thinning; \texttt{burninPerc}, the proportion of burn-in), \texttt{storage} (a list containing numeric values for storing posterior samples for subject- and cluster-specific random effects: \texttt{nGam\_save}, the number of $\gamma$ to be stored; \texttt{storeV}, a vector of three logical values to determine whether all the posterior samples of $\bm{V}_{j}$, for $j=1,\ldots,J$ are to be stored), \texttt{tuning} (a list containing numeric values relevant to tuning parameters for specific updates in Metropolis-Hastings-Green (MHG) algorithm: \texttt{mhProp\_theta\_var}, the variance of proposal density for $\theta$; \texttt{mhProp\_Vg\_var}, the variance of proposal density for $\bm{V}_{j}$ in DPM models; \texttt{mhProp\_alphag\_var}, the variance of proposal density for $\alpha_{g}$ in Weibull models; \texttt{Cg}, a vector of three proportions that determine the sum of probabilities of choosing the birth and the death moves in PEM models (the sum of the three elements should not exceed $0.6$); \texttt{delPertg}, the perturbation parameters in the birth update in PEM models (the values must be between $0$ and $0.5$); \texttt{rj.scheme}: if \texttt{rj.scheme}=1, the birth update will draw the proposal time split from 1:\texttt{sg\_max} and if \texttt{rj.scheme}=2, the birth update will draw the proposal time split from uniquely ordered failure times in the data. For PEM models, additional components are required: \texttt{Kg\_max}, the maximum number of splits allowed at each iteration in MHG algorithm for PEM models; \texttt{time\_lambda1}, \texttt{time\_lambda2}, \texttt{time\_lambda3}, time points at which the posterior distribution of log-hazard functions are calculated. Finally, \texttt{path} indicates the name of directory where the results are saved. For more details and examples, see \cite{khlee2017b}.

\subsection[BayesIDAFT]{\texttt{BayesID\_AFT}} \label{sec:BAFTfc}

The function \texttt{BayesID\_AFT} fits parametric and semi-parametric AFT models for independent semi-competing risks data, given in (\ref{eq:AFTind1})-(\ref{eq:AFTind3}), based on Bayesian inference. Its default structure is given by
\begin{verbatim}
BayesID_AFT(Formula, data, model="LN", hyperParams, startValues, mcmcParams,
path=NULL),
\end{verbatim}
\noindent where \texttt{data}, \texttt{startValues} (auxiliary function \texttt{initiate.startValues\_AFT}), and \texttt{path} are analogous to functions described in previous sections. Here, \texttt{Formula} has a different structure of outcomes, since the AFT model accommodates more complex censoring, such as interval-censoring and/or left-truncation (see Section~\ref{sec:AFTind}). It takes the generic form \texttt{Formula(LT | y1L + y1U | y2L + y2U ~ cov1 | cov2 | cov3)}, where \texttt{LT} represents the left-truncation time, (\texttt{y1L}, \texttt{y1U}) and (\texttt{y2L}, \texttt{y2U}) are the interval-censored times to the non-terminal and terminal events, respectively, and \texttt{cov1}, \texttt{cov2} and \texttt{cov3} are covariates of each linear regression. The \texttt{model} argument specifies whether the baseline survival distribution is parametric (``\texttt{LN}'') or non-parametric (``\texttt{DPM}''). The \texttt{hyperParams} argument defines all model hyperparameters: \texttt{theta} is for hyperparameters ($a^{(\theta)}$ and $b^{(\theta)})$); \texttt{LN} is a list containing numeric vectors, \texttt{LN.ab1}, \texttt{LN.ab2}, \texttt{LN.ab3}, for log-Normal hyperparameters ($a_{g}^{(\sigma)}$, $b_{g}^{(\sigma)}$) with $g \in \left\{1, 2, 3\right\}$; \texttt{DPM} is a list containing numeric vectors, \texttt{DPM.mu1}, \texttt{DPM.mu2}, \texttt{DPM.mu3}, \texttt{DPM.sigSq1}, \texttt{DPM.sigSq2}, \texttt{DPM.sigSq3}, \texttt{DPM.ab1}, \texttt{DPM.ab2}, \texttt{DPM.ab3}, \texttt{Tau.ab1}, \texttt{Tau.ab2}, \texttt{Tau.ab3} for DPM hyperparameters ($\mu_{g0}$, $\sigma_{g0}^{2}$), ($a_{g}^{(\sigma_{gr})}$, $b_{g}^{(\sigma_{gr})}$), and $\tau_{g}$ with $g \in \left\{1, 2, 3\right\}$. The \texttt{mcmcParams} argument sets the information for MCMC sampling: \texttt{run} (see Section \ref{sec:BHRegfc}), \texttt{storage} (\texttt{nGam\_save}; \texttt{nY1\_save}, the number of \texttt{y1} to be stored; \texttt{nY2\_save}, the number of \texttt{y2} to be stored; \texttt{nY1.NA\_save}, the number of \texttt{y1==NA} to be stored), \texttt{tuning} (\texttt{betag.prop.var}, the variance of proposal density for $\bm{\beta}_{g}$; \texttt{mug.prop.var}, the variance of proposal density for $\mu_{g}$; \texttt{zetag.prop.var}, the variance of proposal density for $1/\sigma_{g}^{2}$; \texttt{gamma.prop.var}, the variance of proposal density for $\gamma$).

\subsection{Univariate survival data analysis}

The functions \texttt{FreqSurv\_HReg}, \texttt{BayesSurv\_HReg} and \texttt{BayesSurv\_AFT} provide the same flexibility as functions \texttt{FreqID\_HReg}, \texttt{BayesID\_HReg} and \texttt{BayesID\_AFT}, respectively, but in a univariate context (i.e., a single outcome).

The function \texttt{FreqSurv\_HReg} fits a Weibull PHR model based on maximum likelihood estimation. This model is described by:
\begin{equation}
h(t_{i} \mid \bm{x}_{i}) = \alpha \, \kappa \, t_{i}^{\alpha-1}\exp(\bm{x}_{i}^{\top}\bm{\beta}), ~~ t_{i} > 0. \label{eq:FuniPHR}
\end{equation}

The function \texttt{BayesSurv\_HReg} implements Bayesian PHR models given by:
\begin{equation}
h(t_{ji} \mid \bm{x}_{ji}) = h_{0}(t_{ji})\exp(\bm{x}_{ji}^{\top}\bm{\beta} + V_{j}), ~~ t_{i} > 0, \label{eq:BuniPHR}
\end{equation}
We can adopt either a parametric Weibull or a non-parametric PEM specification for $h_{0}$. Cluster-specific random effects $V_{j}$, $j=1,\ldots,J$, can be assumed to follow a parametric Normal distribution or a non-parametric DPM of Normal distributions.

Finally, the function \texttt{BayesSurv\_AFT} implements Bayesian AFT models expressed by:
\begin{equation}
\log(T_{i}) = \bm{x}_{i}^{\top}\bm{\beta} + \epsilon_{i}, ~~ T_{i} > 0, \label{eq:BuniAFT}
\end{equation}

\noindent where we can adopt either a fully parametric log-Normal or a non-parametric DPM specification for $\epsilon_{i}$ .

\subsection{Summary output} \label{sec:output}

The functions presented in Sections \ref{sec:FHRegfc}, \ref{sec:BHRegfc} and \ref{sec:BAFTfc} return objects of classes \texttt{Freq\_HReg}, \texttt{Bayes\_HReg} and \texttt{Bayes\_AFT}, respectively. Each of these objects represents results from its respective semi-competing risks analysis. These results can be visualized using several R methods, such as \texttt{print}, \texttt{summary}, \texttt{predict}, \texttt{plot}, \texttt{coef}, and \texttt{vcov}.

The function \texttt{print} shows the estimated parameters and, in the Bayesian case, also the MCMC description (number of chains, scans, thinning, and burn-in) and the potential scale reduction factor (PSRF) convergence diagnostic for each model parameter \citep{gelman1992, brooks1998}. If the PSRF is close to 1, a group of chains have mixed well and have converged to a stable distribution. The function \texttt{summary} presents the regression parameters in exponential format (hazard ratios) and the estimated baseline hazard function components. 

Functions \texttt{predict} and \texttt{plot} complement each other. The former uses the fitted model to predict an output of interest (survival or hazard) at a given time interval from new covariates. From the object created by \texttt{predict}, \texttt{plot} displays survival (\texttt{plot.est="Surv"}) or hazard (\texttt{plot.est="Haz"}) functions with their respective $95\%$ confidence/credibility intervals.

Additionally, \textbf{SemiCompRisks} provides the standard functions \texttt{coef} (model coefficients) and \texttt{vcov} (variance-covariance matrix for a fitted frequentist model). For examples with more details, see \cite{khlee2017b}.

\subsection{Simulation of semi-competing risks data}

The function \texttt{simID} simulates semi-competing risks outcomes from independent or cluster-correlated data (for more details of the simulation algorithm, see Appendix~\ref{app:sec1}). The simulation is based on a semi-Markov Weibull PHR modeling and, in the case of the cluster-correlated approach, the cluster-specific random effects follow a MVN distribution. We provide a simulation example of independent semi-competing risks data in Appendix~\ref{app:sec2}.

Analogously, the function \texttt{simSurv} simulates univariate independent/cluster-correlated survival data under a Weibull PHR model with cluster-specific random effects following a Normal distribution.

\subsection{Datasets}

\paragraph{CIBMTR data.} It is composed of $5$ covariates that come from a study of acute GVHD with $9,651$ patients who underwent the first allogeneic hematopoietic cell transplant between January $1999$ and December $2011$ (see Section~\ref{sec:data}).

\paragraph{BMT data.} It refers to a well-known study of bone marrow transplantation for acute leukemia \citep{klein2003}. This data frame contains $137$ patients with $22$ variables and its description can be viewed from the R command \texttt{help(BMT)}.

\section{Illustration: Stem Cell Transplantation data} \label{sec:applic}

In our first example we employ the modeling (\ref{eq:PHRind1})-(\ref{eq:PHRind3}) for independent data, semi-Markov assumption and Weibull baseline hazards. Here, \texttt{Formula} (\texttt{form}) is defined as in Section~\ref{subsec:func}. We fit the model using the function \texttt{FreqID\_HReg}, described in Section~\ref{sec:FHRegfc}, and visualize the results through the function \texttt{summary}:

\subsection{Frequentist analysis} \label{sec:freqanal}

\subsubsection{Independent semi-Markov PHR model with Weibull baseline hazards}

In our first example we employ the modeling (\ref{eq:PHRind1})-(\ref{eq:PHRind3}) for independent data, semi-Markov assumption and Weibull baseline hazards. Here, \texttt{Formula} (\texttt{form}) is defined as in Section~\ref{subsec:func}. We fit the model using the function \texttt{FreqID\_HReg}, described in Section~\ref{sec:FHRegfc}, and visualize the results through the function \texttt{summary}: \small
\begin{verbatim} 
R> fitFreqPHR <- FreqID_HReg(form, data=simCIBMTR, model="semi-Markov")
R> summary(fitFreqPHR)

Analysis of independent semi-competing risks data 
semi-Markov assumption for h3

Hazard ratios:
         beta1   LL  UL beta2   LL  UL beta3   LL  UL
dTypeALL  1.49 1.20 1.8  1.37 1.09 1.7  0.99 0.78 1.3
dTypeCML  1.78 1.41 2.3  0.83 0.64 1.1  1.30 0.99 1.7
dTypeMDS  1.64 1.26 2.1  1.39 1.04 1.9  1.49 1.09 2.0
sexP      0.89 0.79 1.0    NA   NA  NA    NA   NA  NA

Variance of frailties:
      Estimate  LL  UL
theta      7.8 7.3 8.4

Baseline hazard function components:
                   h1-PM   LL    UL  h2-PM     LL     UL  h3-PM     LL     UL
Weibull: log-kappa -6.14 -6.4 -5.90 -11.33 -11.74 -10.93 -6.873 -7.189 -6.557
Weibull: log-alpha  0.15  0.1  0.21   0.86   0.82   0.91  0.022 -0.033  0.077
\end{verbatim} \normalsize

As shown in Section~\ref{sec:output}, \texttt{summary} provides estimates of all model parameters. Using the auxiliary functions \texttt{predict} (default option \texttt{x1new=x2new=x3new=NULL} which corresponds to the baseline specification) and \texttt{plot}, we can graphically visualize the results: \small
\begin{verbatim}
R> pred <- predict(fitFreqPHR, time=seq(0,365,1), tseq=seq(from=0,to=365,by=30))
R> plot(pred, plot.est="Surv")
R> plot(pred, plot.est="Haz")
\end{verbatim} \normalsize

\begin{figure}[!htb]
	\centering
		\includegraphics[width=16cm]{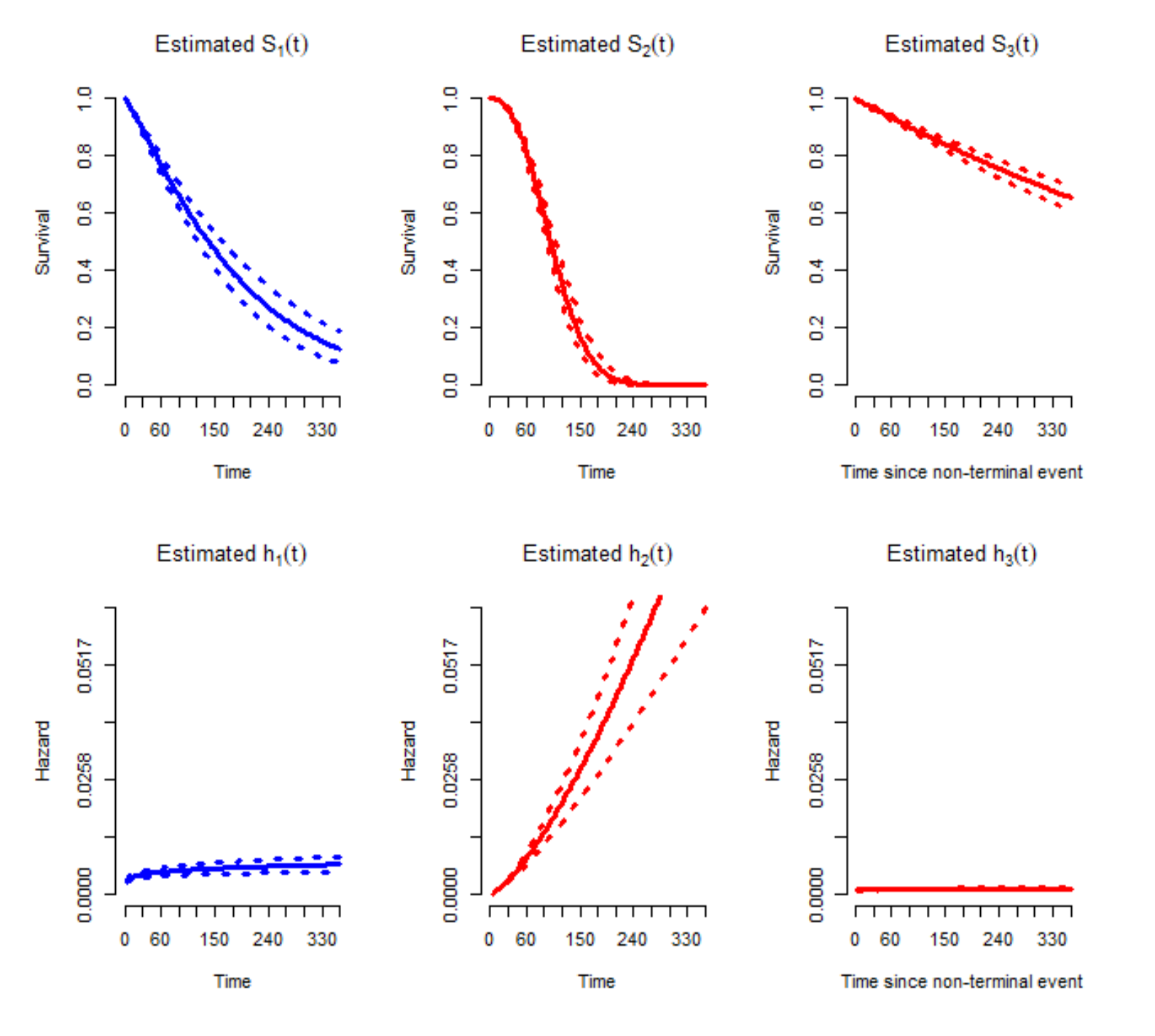}
	\caption{Estimated baseline survival (top) and hazard (bottom) functions from the above analysis.}
	\label{fig:Figure2}	
\end{figure}

Figure~\ref{fig:Figure2} displays estimated baseline survival and hazard functions (solid line) with their corresponding $95\%$ confidence intervals (dotted line).

\subsection{Bayesian analysis} \label{sec:bayesanal}

\subsubsection{Independent semi-Markov PHR model with PEM baseline hazards}

Our second example is also based on the models (\ref{eq:PHRind1})-(\ref{eq:PHRind3}) adopting a semi-Markov assumption for $h_{3}$, but now we use the non-parametric PEM specification for baseline hazard functions. Again, \texttt{Formula} is defined as in Section~\ref{subsec:func}. Here we employ the Bayesian estimation by means of the function \texttt{BayesID\_HReg}, described in Section~\ref{sec:BHRegfc}. The first step is to specify initial values for model parameters through the \texttt{startValues} argument using the auxiliary function \texttt{initiate.startValues\_HReg}: \small
\begin{verbatim}
R> startValues <- initiate.startValues_HReg(form, data=simCIBMTR,
+     model=c("semi-Markov","PEM"), nChain=3)
\end{verbatim} \normalsize

The \texttt{nChain} argument indicates the number of Markov chains that will be used in the MCMC algorithm. Next step is to define all model hyperparameters using the \texttt{hyperParams} argument: \small
\begin{verbatim}
R> hyperParams <- list(theta=c(0.5,0.05), PEM=list(PEM.ab1=c(0.5,0.05),
+     PEM.ab2=c(0.5,0.05), PEM.ab3=c(0.5,0.05), PEM.alpha1=10,
+     PEM.alpha2=10, PEM.alpha3=10))
\end{verbatim} \normalsize

To recall what prior distributions are related to these hyperparameters, see Section~\ref{sec:PHRcluster}. Now we set the MCMC configuration for the \texttt{mcmcParams} argument, more specifically defining the overall run, storage, and tuning parameters for specific updates: \small
\begin{verbatim}
R> sg_max <- c(max(simCIBMTR$time1[simCIBMTR$event1==1]),
+     max(simCIBMTR$time2[simCIBMTR$event1==0 & simCIBMTR$event2==1]),
+     max(simCIBMTR$time2[simCIBMTR$event1==1 & simCIBMTR$event2==1]))

R> mcmcParams <- list(run=list(numReps=5e6, thin=1e3, burninPerc=0.5),
+     storage=list(nGam_save=0, storeV=rep(FALSE,3)),
+     tuning=list(mhProp_theta_var=0.05, Cg=rep(0.2,3), delPertg=rep(0.5,3),
+     rj.scheme=1, Kg_max=rep(50,3), sg_max=sg_max, time_lambda1=seq(1,sg_max[1],1),
+     time_lambda2=seq(1,sg_max[2],1), time_lambda3=seq(1,sg_max[3],1)))
\end{verbatim} \normalsize

As shown above, we set \texttt{sg\_max} to the largest observed failure times for $g \in \left\{1, 2, 3\right\}$. For more details of each item of \texttt{mcmcParams}, see Section~\ref{sec:BHRegfc}.

Given this setup, we fit the PHR model using the function \texttt{BayesID\_HReg}: \small
\begin{verbatim}
R> fitBayesPHR <- BayesID_HReg(form, data=simCIBMTR, model=c("semi-Markov","PEM"), 
+     startValues=startValues, hyperParams=hyperParams, mcmcParams=mcmcParams)
\end{verbatim} \normalsize

We note that, depending on the complexity of the model specification (e.g. if PEM baseline hazards are adopted) and the size of the dataset, despite the functions having been written in C and compiled for R, the MCMC scheme may require a large number of MCMC scans to ensure convergence. As such, some models may take a relatively long time to converge. The example we present below, for example, took 45 hours on a Windows laptop with an Intel(R) Core(TM) i5-3337U 1.80GHz processor, 2 cores, 4 logical processors, 4GB of RAM and 3MB of cache memory to cycle through the 6 millions scans for 3 chains. In lieu of attempting to reproduce the exact results we present here, while readers are of course free to do, Appendix~\ref{app:sec3} provides the code for this same semi-competing risks model and its respective posterior summary, but based on a reduced number of scans of the MCMC scheme (specifically 50,000 scans for 3 chains). Based on the full set of scans, the \texttt{print} method for object returned by \texttt{BayesID\_HReg}, yields: \small
\begin{verbatim}
R> print(fitBayesPHR, digits=2)

Analysis of independent semi-competing risks data 
semi-Markov assumption for h3

Number of chains:     3 
Number of scans:      5e+06 
Thinning:             1000 
Percentage of burnin: 50%

######
Potential Scale Reduction Factor

Variance of frailties, theta:  
 1

Regression coefficients:
         beta1 beta2 beta3
dTypeALL     1     1     1
dTypeCML     1     1     1
dTypeMDS     1     1     1
sexP         1    NA    NA

Baseline hazard function components:

lambda1: summary statistics 
   Min. 1st Qu.  Median    Mean 3rd Qu.    Max. 
   1.00    1.00    1.00    1.00    1.00    1.01 

lambda2: summary statistics 
   Min. 1st Qu.  Median    Mean 3rd Qu.    Max. 
   1.00    1.00    1.00    1.00    1.00    1.01 

lambda3: summary statistics 
   Min. 1st Qu.  Median    Mean 3rd Qu.    Max. 
   1.00    1.00    1.00    1.00    1.01    1.01 

        h1 h2 h3
mu       1  1  1
sigmaSq  1  1  1
K        1  1  1

...
\end{verbatim} \normalsize

Note that all parameters obtained PSRF close to $1$, indicating that the chains have converged well (see Section~\ref{sec:output}). Convergence can also be assessed graphically through a trace plot: \small
\begin{verbatim}
R> plot(fitBayesPHR$chain1$theta.p, type="l", col="red",
+     xlab="iteration", ylab=expression(theta))
R> lines(fitBayesPHR$chain2$theta.p, type="l", col="green")
R> lines(fitBayesPHR$chain3$theta.p, type="l", col="blue")
\end{verbatim} \normalsize
\begin{figure}[!htb]
	\centering
		\includegraphics[width=9cm]{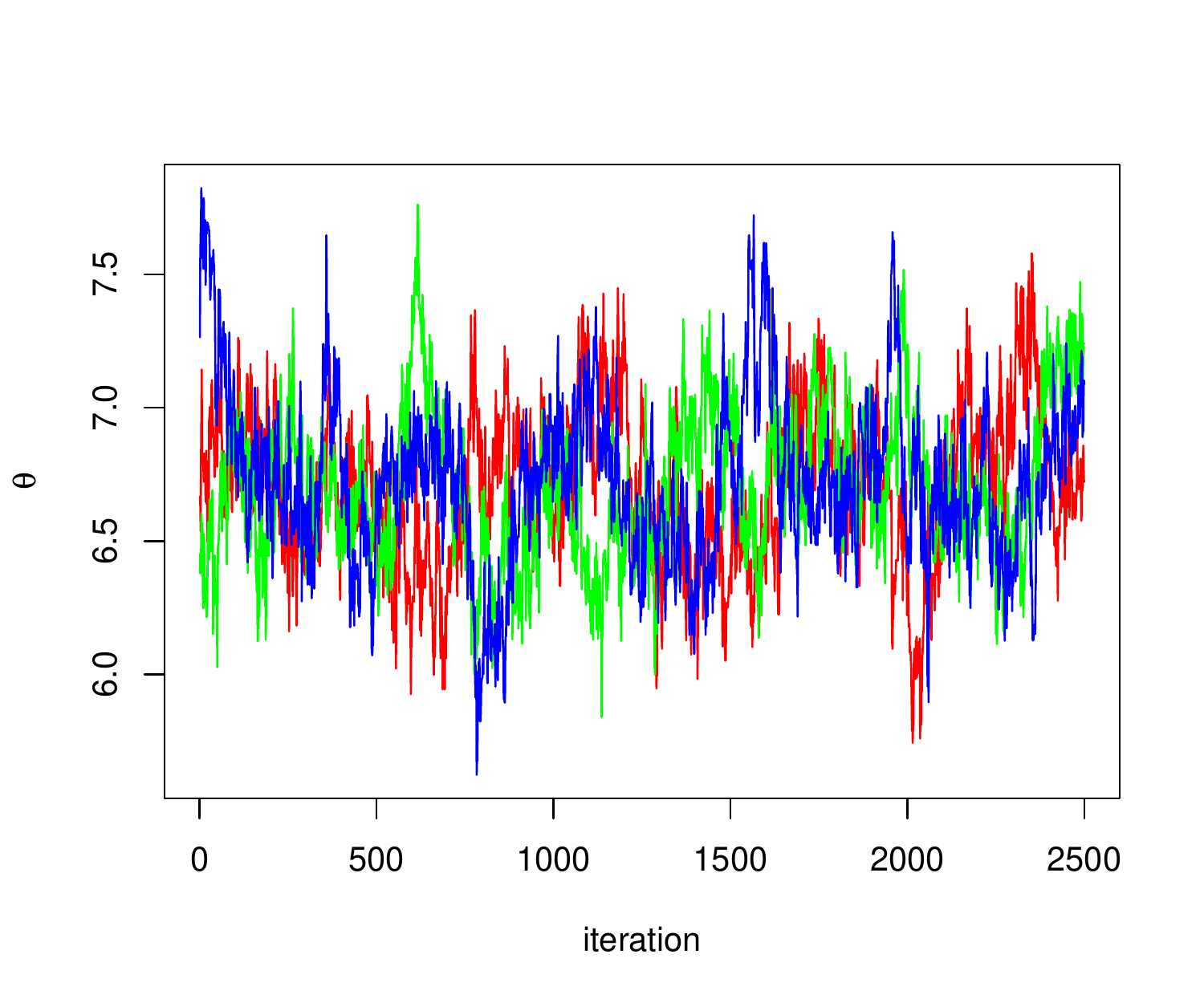}
	\caption{Convergence diagnostic via trace plot of multiple chains.}
	\label{fig:Figure3}
\end{figure}

Figure~\ref{fig:Figure3} shows convergence diagnostic for $\theta$ (subject-specific frailty variance component), where the three chains have mixed and converged to a stable distribution. Any other model parameter could be similarly evaluated. Analogous to the frequentist example, we can also visualize the results through the function \texttt{summary}: \small
\begin{verbatim}
R> summary(fitBayesPHR)

Analysis of independent semi-competing risks data 
semi-Markov assumption for h3

#####

Hazard ratios:
         exp(beta1)   LL  UL exp(beta2)   LL  UL exp(beta3)   LL  UL
dTypeALL       1.45 1.18 1.8       1.32 1.07 1.6       0.99 0.77 1.3
dTypeCML       1.73 1.39 2.2       0.81 0.63 1.0       1.26 0.97 1.6
dTypeMDS       1.61 1.26 2.1       1.36 1.04 1.8       1.45 1.07 1.9
sexP           0.89 0.79 1.0         NA   NA  NA         NA   NA  NA

Variance of frailties:
 theta  LL  UL
   6.7 6.1 7.3

Baseline hazard function components:
        h1-PM     LL UL h2-PM   LL   UL h3-PM     LL   UL
mu      -5.60 -6.003 -5  -5.2 -9.6 -2.3 -6.75 -7.048 -6.5
sigmaSq  0.21  0.027  2   7.2  2.5 24.3  0.13  0.018  2.4
K       11.00  5.000 17  15.0 11.0 21.0 10.00  4.000 17.0
\end{verbatim} \normalsize

Here we provide estimates of all model parameters with their respective $95\%$ credible intervals.

\subsubsection{Independent AFT model with log-Normal baseline survival distribution}

Our last example is based on AFT models (\ref{eq:AFTind1})-(\ref{eq:AFTind3}) adopting a semi-Markov assumption for $h_{3}$ and the parametric log-Normal specification for baseline survival distributions. Here we apply the Bayesian framework via function \texttt{BayesID\_AFT}. As pointed out in Section~\ref{sec:BAFTfc}, \texttt{Formula} argument for AFT models takes a specific form: \small
\begin{verbatim}
R> simCIBMTR$LT <- rep(0,dim(simCIBMTR)[1])
R> simCIBMTR$y1L <- simCIBMTR$y1U <- simCIBMTR[,1]
R> simCIBMTR$y1U[which(simCIBMTR[,2]==0)] <- Inf
R> simCIBMTR$y2L <- simCIBMTR$y2U <- simCIBMTR[,3]
R> simCIBMTR$y2U[which(simCIBMTR[,4]==0)] <- Inf

R> formAFT <- Formula(LT | y1L + y1U | y2L + y2U ~ dTypeALL + dTypeCML + dTypeMDS +
+     sexP | dTypeALL + dTypeCML + dTypeMDS | dTypeALL + dTypeCML + dTypeMDS)
\end{verbatim} \normalsize

Recall that \texttt{LT} represents the left-truncation time, and (\texttt{y1L}, \texttt{y1U}) and (\texttt{y2L}, \texttt{y2U}) are the interval-censored times to the non-terminal and terminal events, respectively. Next step is to set the initial values for model parameters through the \texttt{startValues} argument, but now using the auxiliary function \texttt{initiate.startValues\_AFT}: \small
\begin{verbatim}
R> startValues <- initiate.startValues_AFT(formAFT, data=simCIBMTR,
+     model="LN", nChain=3)
\end{verbatim} \normalsize

Again, we considered three Markov chains (\texttt{nChain}=3). Using the \texttt{hyperParams} argument we specify all model hyperparameters: \small
\begin{verbatim}
R> hyperParams <- list(theta=c(0.5,0.05), LN=list(LN.ab1=c(0.5,0.05),
+     LN.ab2=c(0.5,0.05), LN.ab3=c(0.5,0.05)))
\end{verbatim} \normalsize

Each pair of hyperparameters defines shape and scale of an inverse Gamma prior distribution (see Section~\ref{sec:AFTind}). Similar to the previous example, we must specify overall run, storage, and tuning parameters for specific updates through the \texttt{mcmcParams} argument: \small
\begin{verbatim}
R> mcmcParams <- list(run=list(numReps=5e6, thin=1e3, burninPerc=0.5),
+     storage=list(nGam_save=0, nY1_save=0, nY2_save=0, nY1.NA_save=0),
+     tuning=list(betag.prop.var=rep(0.01,3), mug.prop.var=rep(0.01,3),
+     zetag.prop.var=rep(0.01,3), gamma.prop.var=0.01))
\end{verbatim} \normalsize

Analogous to the previous Bayesian model, a large number of scans are also required here to achieve the convergence of the Markov chains. Again, for a quickly reproducible example, the code for the AFT model with simplified MCMC setting is provided in Appendix~\ref{app:sec3}. For more details of each item of \texttt{mcmcParams}, see Section~\ref{sec:BAFTfc}. Finally, we fit the AFT model using the function \texttt{BayesID\_AFT} and analyze the convergence of each parameter through the function \texttt{print}: \small
\begin{verbatim}
R> fitBayesAFT <- BayesID_AFT(formAFT, data=simCIBMTR, model="LN",
+     startValues=startValues, hyperParams=hyperParams, mcmcParams=mcmcParams)
R> print(fitBayesAFT, digits=2)

Analysis of independent semi-competing risks data 

Number of chains:     3 
Number of scans:      5e+06 
Thinning:             1000 
Percentage of burnin: 50%

######
Potential Scale Reduction Factor

Variance of frailties, theta:  
 1

Regression coefficients:
         beta1 beta2 beta3
dTypeALL     1     1     1
dTypeCML     1     1     1
dTypeMDS     1     1     1
sexP         1    NA    NA

Baseline survival function components:
        g=1 g=2 g=3
mu        1 1.1   1
sigmaSq   1 1.0   1

...
\end{verbatim} \normalsize

Again, the PSRF for each parameter indicates the convergence. As a last step, we visualize the estimate of each parameter and their respective $95\%$ credible intervals through the function \texttt{summary}: \small
\begin{verbatim}
R> summary(fitBayesAFT)
 
Analysis of independent semi-competing risks data 

#####

Acceleration factors:
         exp(beta1)   LL   UL exp(beta2)   LL  UL exp(beta3)   LL  UL
dTypeALL       0.68 0.55 0.85       0.95 0.86 1.0       1.09 0.86 1.4
dTypeCML       0.53 0.42 0.67       1.27 1.11 1.4       0.91 0.71 1.2
dTypeMDS       0.58 0.44 0.76       0.89 0.78 1.0       0.78 0.59 1.0
sexP           1.16 0.99 1.36         NA   NA  NA         NA   NA  NA

Variance of frailties:
 theta  LL  UL
   2.6 2.5 2.8

Baseline survival function components:
                    g=1: PM  LL  UL g=2: PM    LL    UL g=3: PM  LL  UL
log-Normal: mu          8.2 8.0 8.4   6.274 6.226 6.323     6.5 6.4 6.7
log-Normal: sigmaSq     7.1 6.4 8.0   0.014 0.006 0.038     1.7 1.5 2.0
\end{verbatim} \normalsize

\section{Discussion} \label{sec:disc}

This paper discusses the implementation of a comprehensive R package \textbf{SemiCompRisks} for the analyses of independent/cluster-correlated semi-competing risks data. The package allows to fit parametric or semi-parametric models based on either accelerated failure time or proportional hazards regression approach. It is also flexible in that one can adopt either a Markov or semi-Markov specification for terminal event following non-terminal event. The estimation and inference are mostly based on the Bayesian paradigm, but parametric PHR models can also be fitted using the maximum likelihood estimation. Users can easily obtain numerical and graphical presentation of model fits using R methods, as illustrated in the stem cell transplantation example in Section~\ref{sec:applic}. In addition, the package provides functions for performing univariate survival analysis. We would also like to emphasize that the vignette documentation \citep{khlee2017b} provides a list of detailed examples applying each of the implemented models in the package.

\textbf{SemiCompRisks} provides researchers with valid and practical analysis tools for semi-competing risks data. The application examples in this paper were run using version v3.0 of the package, available from the CRAN at \url{https://cran.r-project.org/package=SemiCompRisks}. We plan to constantly update the package to incorporate more functionality and flexibility to the models for semi-competing risks analysis. 

\section*{Acknowledgments}

Funding for this work was provided by National Institutes of Health grants R01 CA181360-01. The authors also gratefully acknowledge the CIBMTR (grant U24-CA076518) for providing the covariates of the illustrative example.

\bibliographystyle{plainnat}
\bibliography{references}

\newpage

\appendix

\section{Simulation algorithm for semi-competing risks data} \label{app:sec1}

The \textbf{SemiCompRisks} package contains a function, \texttt{simID}, for simulating independent or cluster-correlated semi-competing risks data. In this section, we provide the details on the simulation algorithm used in \texttt{simID} for generating cluster-correlated semi-competing risks data based on a parametric Weibull-MVN semi-Markov illness-death model, as presented in Section \ref{sec:PHRcluster}, where the baseline hazard functions are defined as $h_{0g}(t)=\alpha_{g} \, \kappa_{g} \, t^{\alpha_{g}-1}$, for $g \in \left\{1, 2, 3\right\}$. The step by step algorithm is given as follows:

\begin{enumerate}
	\item Generate $\bm{V}_{j}=(V_{j1}, V_{j2}, V_{j3})^{\top}$ from a MVN($\bm{0}$, $\Sigma_{V}$), for $j=1,\ldots,J$.
	\item For each $j$, repeat the following steps for $i=1,\ldots,n_{j}$.
	\begin{enumerate}[a)]
		\item Generate $\gamma_{ji}$ from a Gamma($\theta^{-1}$, $\theta^{-1}$).
		\item Calculate $\eta_{jig}=\log(\gamma_{ji})+\bm{x}_{jig}^{\top}\bm{\beta}_{g} + V_{jg}$, for $g \in \left\{1, 2, 3\right\}$.
		\item Generate $t_{1}^{\ast}$ from a Weibull($\alpha_{1}$, $\kappa_{1} \, e^{\eta_{ji1}})$ and $t_{2}^{\ast}$ from a Weibull($\alpha_{2}$, $\kappa_{2} \, e^{\eta_{ji2}})$.
		\begin{itemize}
			\item If $t_{1}^{\ast} \leq t_{2}^{\ast}$, generate $t^{\ast}$ from a Weibull($\alpha_{3}$, $\kappa_{3} \, e^{\eta_{ji3}})$ and set $t_{ji1}=t_{1}^{\ast}$, $t_{ji2}=t_{1}^{\ast}+t^{\ast}$.
			\item Otherwise, set $t_{ji1}=\infty$, $t_{ji2}=t_{2}^{\ast}$.
		\end{itemize}
		\item Generate a censoring time $c_{ji}$ from Uniform($c_{L}$, $c_{U}$).
		\item Set the observed outcome information (\texttt{time1}, \texttt{time2}, \texttt{event1}, \texttt{event2}) as follows:
		\begin{itemize}
			\item ($t_{ji1}$, $t_{ji2}$, 1, 1), if $t_{ji1}<t_{ji2}<c_{ji}$.
			\item ($t_{ji1}$, $c_{ji}$, 1, 0), if $t_{ji1}<c_{ji}<t_{ji2}$.
			\item ($t_{ji2}$, $t_{ji2}$, 0, 1), if $t_{ji1}=\infty$ and $t_{ji2}<c_{ji}$.
			\item ($c_{ji}$, $c_{ji}$, 0, 0), if $t_{ji1}>c_{ji}$ and $t_{ji2}>c_{ji}$.
		\end{itemize}		
	\end{enumerate}	
\end{enumerate}

We note that the function \texttt{simID} is flexible in that one can set the $\theta$ argument as zero (\texttt{theta.true}=0) to simulate the data under the model without the subject-specific shared frailty term ($\gamma_{ji}$), which is analogous to the model proposed by \cite{liquet2012}. One can generate independent semi-competing risks data outlined in Section \ref{sec:PHRind} by setting the \texttt{id} and $\Sigma_{V}$ arguments as nulls (\texttt{cluster}=\texttt{NULL} and \texttt{SimgaV.true}=\texttt{NULL}).

\section{Simulating outcomes using CIBMTR covariates} \label{app:sec2}

The true values of model parameters are set to estimates obtained by fitting a semi-Markov Weibull PHR model to the original CIBMTR data. \small
\begin{verbatim}
R> data(CIBMTR_Params)
R> beta1.true <- CIBMTR_Params$beta1.true 
R> beta2.true <- CIBMTR_Params$beta2.true
R> beta3.true <- CIBMTR_Params$beta3.true
R> alpha1.true <- CIBMTR_Params$alpha1.true
R> alpha2.true <- CIBMTR_Params$alpha2.true
R> alpha3.true <- CIBMTR_Params$alpha3.true
R> kappa1.true <- CIBMTR_Params$kappa1.true
R> kappa2.true <- CIBMTR_Params$kappa2.true
R> kappa3.true <- CIBMTR_Params$kappa3.true
R> theta.true <- CIBMTR_Params$theta.true
R> cens <- c(365, 365)
\end{verbatim} \normalsize

The next step is to define the covariates matrices and then simulate outcomes using the \texttt{simID} function, available in the \textbf{SemiCompRisks} package. \small
\begin{verbatim}
R> data(CIBMTR)
# Sex (M: reference category)
R> CIBMTR$sexP <- as.numeric(CIBMTR$sexP)-1

# Age (LessThan10: reference category)
R> CIBMTR$ageP20to29 <- as.numeric(CIBMTR$ageP=="20to29")
R> CIBMTR$ageP30to39 <- as.numeric(CIBMTR$ageP=="30to39")
R> CIBMTR$ageP40to49 <- as.numeric(CIBMTR$ageP=="40to49")
R> CIBMTR$ageP50to59 <- as.numeric(CIBMTR$ageP=="50to59")
R> CIBMTR$ageP60plus <- as.numeric(CIBMTR$ageP=="60plus")

# Disease type (AML: reference category)
R> CIBMTR$dTypeALL <- as.numeric(CIBMTR$dType=="ALL")
R> CIBMTR$dTypeCML <- as.numeric(CIBMTR$dType=="CML")
R> CIBMTR$dTypeMDS <- as.numeric(CIBMTR$dType=="MDS")

# Disease status (Early: reference category)
R> CIBMTR$dStatusInt <- as.numeric(CIBMTR$dStatus=="Int")
R> CIBMTR$dStatusAdv <- as.numeric(CIBMTR$dStatus=="Adv")

# HLA compatibility (HLA_Id_Sib: reference category)
R> CIBMTR$donorGrp8_8 <- as.numeric(CIBMTR$donorGrp=="8_8")
R> CIBMTR$donorGrp7_8 <- as.numeric(CIBMTR$donorGrp=="7_8")

# Covariate matrix
R> x1 <- CIBMTR[,c("sexP", "ageP20to29", "ageP30to39", "ageP40to49",
+     "ageP50to59", "ageP60plus", "dTypeALL", "dTypeCML", "dTypeMDS",
+     "dStatusInt", "dStatusAdv", "donorGrp8_8", "donorGrp7_8")]

R> x2 <- CIBMTR[,c("sexP", "ageP20to29", "ageP30to39", "ageP40to49",
+     "ageP50to59", "ageP60plus", "dTypeALL", "dTypeCML", "dTypeMDS",
+     "dStatusInt", "dStatusAdv", "donorGrp8_8", "donorGrp7_8")]

R> x3 <- CIBMTR[,c("sexP", "ageP20to29", "ageP30to39", "ageP40to49",
+     "ageP50to59", "ageP60plus", "dTypeALL", "dTypeCML", "dTypeMDS",
+     "dStatusInt", "dStatusAdv", "donorGrp8_8", "donorGrp7_8")]

R> set.seed(1405)
R> simOutcomes <- simID(id=NULL, x1=x1, x2=x2, x3=x3,
+     beta1.true, beta2.true, beta3.true, alpha1.true, alpha2.true, alpha3.true,
+     kappa1.true, kappa2.true, kappa3.true, theta.true, SigmaV.true=NULL, cens)

R> names(simOutcomes) <- c("time1", "event1", "time2", "event2")
R> simCIBMTR <- cbind(simOutcomes, CIBMTR[,c("sexP", "ageP20to29", "ageP30to39", 
+    "ageP40to49", "ageP50to59", "ageP60plus", "dTypeALL", "dTypeCML", "dTypeMDS", 
+    "dStatusInt", "dStatusAdv", "donorGrp8_8", "donorGrp7_8")])
\end{verbatim} \normalsize

\section{Code for illustrative Bayesian examples} \label{app:sec3}

In order to encourage the reproducibility of the results obtained through our R package in a reasonable computational time, Bayesian analyses contained in Section~\ref{sec:bayesanal} are illustrated below using a reduced number of scans (\texttt{numReps}) and extent of thinning (\texttt{thin}). Given the complexity of these Bayesian models, the reduction of scans/thinning results in non-convergence of the Markov chains, but at least it is possible to reproduce the results quickly.

\subsection{Independent semi-Markov PHR model with PEM baseline hazards}

\small
\begin{verbatim}
R> form <- Formula(time1 + event1 | time2 + event2 ~ dTypeALL + dTypeCML + 
+     dTypeMDS + sexP | dTypeALL + dTypeCML + dTypeMDS | dTypeALL +
+     dTypeCML + dTypeMDS)

R> startValues <- initiate.startValues_HReg(form, data=simCIBMTR,
+     model=c("semi-Markov","PEM"), nChain=3)

R> hyperParams <- list(theta=c(0.5,0.05), PEM=list(PEM.ab1=c(0.5,0.05),
+     PEM.ab2=c(0.5,0.05), PEM.ab3=c(0.5,0.05), PEM.alpha1=10,
+     PEM.alpha2=10, PEM.alpha3=10))

R> sg_max <- c(max(simCIBMTR$time1[simCIBMTR$event1==1]),
+     max(simCIBMTR$time2[simCIBMTR$event1==0 & simCIBMTR$event2==1]),
+     max(simCIBMTR$time2[simCIBMTR$event1==1 & simCIBMTR$event2==1]))

R> mcmcParams <- list(run=list(numReps=5e4, thin=5e1, burninPerc=0.5),
+     storage=list(nGam_save=0, storeV=rep(FALSE,3)),
+     tuning=list(mhProp_theta_var=0.05, Cg=rep(0.2,3), delPertg=rep(0.5,3),
+     rj.scheme=1, Kg_max=rep(50,3), sg_max=sg_max, time_lambda1=seq(1,sg_max[1],1),
+     time_lambda2=seq(1,sg_max[2],1), time_lambda3=seq(1,sg_max[3],1)))

R> fitBayesPHR <- BayesID_HReg(form, data=simCIBMTR, model=c("semi-Markov","PEM"), 
+     startValues=startValues, hyperParams=hyperParams, mcmcParams=mcmcParams)
R> print(fitBayesPHR, digits=2)

Analysis of independent semi-competing risks data 
semi-Markov assumption for h3

Number of chains:     3 
Number of scans:      50000 
Thinning:             50 
Percentage of burnin: 50%

######
Potential Scale Reduction Factor

Variance of frailties, theta:    
 3.4

Regression coefficients:
         beta1 beta2 beta3
dTypeALL   1.8   1.6   1.8
dTypeCML   1.9   1.8   2.1
dTypeMDS   1.8   2.0   2.0
sexP       1.1    NA    NA

Baseline hazard function components:

lambda1: summary statistics 
   Min. 1st Qu.  Median    Mean 3rd Qu.    Max. 
    1.1     2.9     3.2     3.2     3.6     4.3 

lambda2: summary statistics 
   Min. 1st Qu.  Median    Mean 3rd Qu.    Max. 
    1.0     3.3     4.0     3.7     4.5     5.8 

lambda3: summary statistics 
   Min. 1st Qu.  Median    Mean 3rd Qu.    Max. 
    1.8     2.2     2.5     2.5     2.7     3.2 

         h1  h2  h3
mu      1.3 2.2 1.5
sigmaSq 1.0 1.1 1.0
K       1.1 1.0 1.0

######
Estimates

Variance of frailties, theta:
 Estimate  SD  LL UL
       10 0.8 9.4 12

Regression coefficients:
         Estimate   SD   LL   UL
dTypeALL     0.10 0.19 0.82 1.59
dTypeCML     0.31 0.21 0.95 2.09
dTypeMDS     0.19 0.20 0.89 1.83
sexP        -0.14 0.06 0.79 0.99
dTypeALL    -0.03 0.21 0.69 1.41
dTypeCML    -0.52 0.25 0.40 0.99
dTypeMDS    -0.04 0.21 0.69 1.48
dTypeALL    -0.34 0.24 0.47 1.13
dTypeCML     0.00 0.23 0.67 1.56
dTypeMDS     0.15 0.26 0.71 1.82
\end{verbatim} \normalsize

\subsection{Independent AFT model with log-Normal baseline survival distribution}

\small
\begin{verbatim}
R> simCIBMTR$LT <- rep(0,dim(simCIBMTR)[1])
R> simCIBMTR$y1L <- simCIBMTR$y1U <- simCIBMTR[,1]
R> simCIBMTR$y1U[which(simCIBMTR[,2]==0)] <- Inf
R> simCIBMTR$y2L <- simCIBMTR$y2U <- simCIBMTR[,3]
R> simCIBMTR$y2U[which(simCIBMTR[,4]==0)] <- Inf

R> formAFT <- Formula(LT | y1L + y1U | y2L + y2U ~ dTypeALL + dTypeCML + dTypeMDS +
+     sexP | dTypeALL + dTypeCML + dTypeMDS | dTypeALL + dTypeCML + dTypeMDS)

R> startValues <- initiate.startValues_AFT(formAFT, data=simCIBMTR,
+     model="LN", nChain=3)

R> hyperParams <- list(theta=c(0.5,0.05), LN=list(LN.ab1=c(0.5,0.05),
+     LN.ab2=c(0.5,0.05), LN.ab3=c(0.5,0.05)))

R> mcmcParams <- list(run=list(numReps=5e4, thin=5e1, burninPerc=0.5),
+     storage=list(nGam_save=0, nY1_save=0, nY2_save=0, nY1.NA_save=0),
+     tuning=list(betag.prop.var=rep(0.01,3), mug.prop.var=rep(0.01,3),
+     zetag.prop.var=rep(0.01,3), gamma.prop.var=0.01))

R> fitBayesAFT <- BayesID_AFT(formAFT, data=simCIBMTR, model="LN",
+     startValues=startValues, hyperParams=hyperParams, mcmcParams=mcmcParams)
R> print(fitBayesAFT, digits=2)

Analysis of independent semi-competing risks data 

Number of chains:     3 
Number of scans:      50000 
Thinning:             50 
Percentage of burnin: 50%

######
Potential Scale Reduction Factor

Variance of frailties, theta:    
 1.7

Regression coefficients:
         beta1 beta2 beta3
dTypeALL     1   1.1   1.1
dTypeCML     1   1.0   1.0
dTypeMDS     1   1.1   1.0
sexP         1    NA    NA

Baseline survival function components:
        g=1 g=2 g=3
mu      1.2 1.1 1.1
sigmaSq 1.4 1.7 1.2

######
Estimates

Variance of frailties, theta:
 Estimate   SD  LL UL
      1.7 0.16 1.4  2

Regression coefficients:
         Estimate   SD   LL   UL
dTypeALL    -0.42 0.13 0.52 0.83
dTypeCML    -0.73 0.13 0.37 0.62
dTypeMDS    -0.60 0.14 0.42 0.72
sexP         0.14 0.09 0.95 1.37
dTypeALL    -0.02 0.06 0.87 1.09
dTypeCML     0.30 0.06 1.20 1.54
dTypeMDS    -0.10 0.07 0.78 1.01
dTypeALL     0.04 0.13 0.81 1.31
dTypeCML    -0.14 0.12 0.67 1.10
dTypeMDS    -0.23 0.14 0.61 1.03
\end{verbatim}

\end{document}